\documentclass[12pt]{article}
\usepackage{epsfig}
\newcommand{\be}{\begin{equation}}
\newcommand{\ee}{\end{equation}}
\newcommand{\bdm}{\begin{displaymath}}
\newcommand{\edm}{\end{displaymath}}
\renewcommand{\thefootnote}{\fnsymbol{footnote}}
\def\simlt{\mathrel{\lower2.5pt\vbox{\lineskip=0pt\baselineskip=0pt
           \hbox{$<$}\hbox{$\sim$}}}}
\def\simgt{\mathrel{\lower2.5pt\vbox{\lineskip=0pt\baselineskip=0pt
           \hbox{$>$}\hbox{$\sim$}}}}
\newcommand{\ls}[1]
   {\dimen0=\fontdimen6\the\font
    \lineskip=#1\dimen0
    \advance\lineskip.5\fontdimen5\the\font
    \advance\lineskip-\dimen0
    \lineskiplimit=.9\lineskip
    \baselineskip=\lineskip
    \advance\baselineskip\dimen0
    \normallineskip\lineskip
    \normallineskiplimit\lineskiplimit
    \normalbaselineskip\baselineskip
    \ignorespaces}

\catcode`@=11
\newcount\@tempcntc
\def\@citex[#1]#2{\if@filesw\immediate\write\@auxout{\string\citation{#2}}\fi
  \@tempcnta\z@\@tempcntb\m@ne\def\@citea{}\@cite{\@for\@citeb:=#2\do
    {\@ifundefined
       {b@\@citeb}{\@citeo\@tempcntb\m@ne\@citea\def\@citea{,}{\bf ?}\@warning
       {Citation `\@citeb' on page \thepage \space undefined}}%
    {\setbox\z@\hbox{\global\@tempcntc0\csname b@\@citeb\endcsname\relax}%
     \ifnum\@tempcntc=\z@ \@citeo\@tempcntb\m@ne
       \@citea\def\@citea{,}\hbox{\csname b@\@citeb\endcsname}%
     \else
      \advance\@tempcntb\@ne
      \ifnum\@tempcntb=\@tempcntc
      \else\advance\@tempcntb\m@ne\@citeo
      \@tempcnta\@tempcntc\@tempcntb\@tempcntc\fi\fi}}\@citeo}{#1}}
\def\@citeo{\ifnum\@tempcnta>\@tempcntb\else\@citea\def\@citea{,}%
  \ifnum\@tempcnta=\@tempcntb\the\@tempcnta\else
   {\advance\@tempcnta\@ne\ifnum\@tempcnta=\@tempcntb \else \def\@citea{--}\fi
    \advance\@tempcnta\m@ne\the\@tempcnta\@citea\the\@tempcntb}\fi\fi}
\catcode`@=12

\begin{document}
\setcounter{footnote}{1}
\begin{flushright}
\end{flushright}
\vspace{7mm}
\begin{center}
\Large{{\bf 
Constraints on the Bulk Standard Model in the Randall-Sundrum Scenario
}}
\end{center}
\vspace{5mm}
\begin{center}
{\large\bf Gustavo Burdman\\
\vspace{0.3cm}
{\normalsize\it 
Theoretical Physics  Group, 
Lawrence Berkeley National Laboratory  
Berkeley, CA 94720}
}
\end{center}
\vspace{0.50cm}
\thispagestyle{empty}
\begin{abstract}
We derive constraints on the Randall-Sundrum scenario with the standard model 
fields in the bulk. These result from tree level effects associated with 
the deformation of the zero mode wave-functions of the $W$ and the $Z$ once electroweak 
symmetry is broken. 
Recently Cs\'{a}ki, Erlich and Terning pointed out that this implies large contributions
to electroweak oblique parameters. Here we 
find that when fermions are allowed in the bulk
the couplings of the $W$ and the $Z$ to zero-mode fermions are also affected.
We perform a fit to electroweak observables assuming universal bulk fermion masses and
including all effects and find constraints that are considerably 
stronger than for the case with fermions localized in the low energy boundary.
These put the lowest Kaluza-Klein excitation out of reach of
the Large Hadron Collider. 
We then relax the universality assumption and study the effects of flavor violation 
in the bulk and its possible signatures.

\end{abstract}
\newpage

\renewcommand{\thefootnote}{\arabic{footnote}}
\setcounter{footnote}{0}
\setcounter{page}{1}

\section{Introduction}

Theories with large extra dimensions have been recently introduced as an alternative
framework to solve the hierarchy problem~\cite{nima1}.  It is assumed that the geometry
is factorizable and results in a product of Minkowski space with $n$ compact dimensions.
In this scenario, gravity propagates in the extra dimensions so that the strength of
its coupling to matter confined in our four dimensional world is determined by the scale 
$M_{\rm P}^2 = M^{n+2} V_n$, with $M$ the fundamental scale of gravity and $V_n$
the volume of the extra dimensions. In this way, the hierarchy between $M_{\rm P}$ 
and the weak scale results from the volume suppression, and the truly fundamental 
scale $M$ can be of the order of $1$~TeV. 

An alternative scenario by Randall and Sundrum involves the use 
of a non-factorizable geometry in five dimensions~\cite{rs1}.
The metric depends on the five dimensional coordinate $y$ and is given by 
\begin{equation}
ds^2 = e^{-2\sigma(y)} \eta_{\mu\nu} dx^\mu dx^\nu - dy^2~,
\label{metric}
\end{equation} 
where $x^\mu$ are the four dimensional coordinates, $\sigma(y) = k |y|$, with 
$k\sim M_P$ characterizing the curvature scale. The extra dimension is compactified
on an orbifold $S_1/Z_2$ of radius $r$ so that the bulk is a slice of ${\rm AdS}_5$
space between two four-dimensional boundaries. The metric on these boundaries generates
two effective scales: $M_P$ and $M_P e^{-k\pi r}$. In this way, values of 
$r$ not much larger than the Planck length can be used in order to generate
a scale $\Lambda_r\simeq M_Pe^{-k\pi r}\simeq~{\rm O(TeV)}$, i.e. $kr\simeq (11-12)$, 
for generating the TeV on one of the boundaries. 

In the original RS scenario, only gravity was allowed to propagate in the bulk, 
with the Standard Model (SM) fields confined to one of the boundaries. 
The inclusion of matter and gauge fields in the bulk has been extensively treated in the
literature~\cite{bulk1,chang,gn,gp,huber1,dhr1,dhr2}. 
In this paper we are interested in examining the situation when the SM fields 
are allowed to propagate in the bulk. The exception in this so called bulk SM, is the 
Higgs field which must be localized on the TeV boundary in order for 
the $W$ and the $Z$ gauge bosons to get their observed 
masses~\cite{chang}. As it was first noted in Ref.~\cite{huber1}  
the wave functions of the $W$ and $Z$ acquire a dependence on the fifth dimensional
coordinate due to the Higgs vacuum expectation value (VEV). 
Recently Cs\'{a}ki, Erlich and Terning~\cite{cet} have studied 
the effects that result from this deformation of the zero modes
in a scenario with only gauge fields in the bulk. 
They found large contributions to the oblique parameters $S$ and 
$T$ and the bound $\Lambda_r>11$~TeV, which is slightly tighter than 
the ones previously obtained~\cite{dhr1}. 
In this paper we consider a scenario where all or at least part of the fermions
can propagate in the bulk. It is generally believed that this relaxes the 
bounds on $\Lambda_r$ since the couplings of Kaluza-Klein excitations of gauge 
bosons to zero-mode fermions are not as strong as when fermion are confined to the 
TeV boundary.  
Here we show that there are additional effects resulting from the modified 
couplings of $W$ and $Z$ to the SM fermions that propagate in the bulk. 
Even when we consider  these to be flavor universal, they result in non-oblique contributions to 
electroweak observables and in {\em stronger}   
constraints on $\Lambda_r$ than the ones obtained with confined fermions. 
If flavor breaking in the bulk is allowed, then there are additional effects
in flavor changing neutral processes.

In the next Section we review the bulk SM and the existing bounds on 
the induced low energy scale $\Lambda_r$. In Section~\ref{bounds} we 
obtain the  bounds on $\Lambda_r$ coming from the deviation in the tree level 
couplings of fermions to $W$ and $Z$, adding this effect to the contributions
discussed in Ref.~\cite{cet}. We derive these new constraints by making the 
simplifying assumption that the effects on the fermion couplings are flavor universal. 
In Section~\ref{flavor} we study the effects of flavor violation in the bulk.
Finally, in Section~\ref{conclusions} we conclude.

\section{The Bulk Standard Model}
\label{bsm}

The five-dimensional action for 
bulk gauge fields is given by~\cite{bulk1,chang}:
\begin{equation}
S_{A} = -\frac{1}{4}  \int d^4x dy \sqrt{-g} F^{MN}F_{MN}~,
\label{bulkgauge}
\end{equation}
where  $g={\rm det}(g_{MN})=e^{-4\sigma(y)}$ and 
capital latin letters denote five dimensional indexes.
The field strength
is in general written as
\begin{equation}
F_{MN} \equiv \partial_MA_N - \partial_NA_M +ig_5[A_M,A_N]~.
\label{fmn}
\end{equation}
We make the  choice of gauge $A_y=0$. The Kaluza-Klein decomposition is
given by 
\begin{equation}
A_\mu(x,y) = \frac{1}{\sqrt{2\pi r}}\sum_{n=0}^{\infty} A_\mu^{(n)}(x) \chi^{(n)}(y)~,
\label{kkgauge}
\end{equation}
Thus, 
the wave function of the gauge boson in the fifth dimension
$\chi^{(n)}(y)$ obeys the differential equation
\begin{equation}
-\partial_y\left(e^{-2\sigma}\partial_y\chi^{(n)}\right) = m_n^2 \chi^{(n)}~.
\label{diffeq}
\end{equation}
The solutions satisfy the normalization condition:
\begin{equation}
\frac{1}{2\pi r}{\int}_{\!\!\!-\pi r}^{\pi r} dy~\chi^{(m)} 
\chi^{(n)} = \delta_{mn}~,
\label{normal}
\end{equation}
and are 
\begin{equation}
\chi^{(n)}(y) = \frac{e^{\sigma}}{N_n}\left[J_1(\frac{m_n}{\kappa}e^{\sigma}) 
+\alpha_n Y_1(\frac{m_n}{\kappa}e^{\sigma})\right]~.
\label{chins}
\end{equation}
In eqn.~(\ref{chins}), $N_n$ is the normalization constant derived from (\ref{normal}), 
and 
\begin{equation}
\alpha_n = ~-\frac{J_0(m_n/k)}{Y_0(m_n/k)}~, 
\label{alpha}
\end{equation}
where we defined 
\begin{equation}
x_n\equiv \frac{m_n}{k}\; e^{kr\pi} ~=~ \frac{m_n}{\Lambda_r}~,
\label{xndef}
\end{equation}
i.e. the mass of the KK excitation in units of the generated low energy scale.
The zero mode is flat in the extra dimension: $\chi^{(0)}(y)=1$. 
Imposing continuity at $y=(0,\pi r)$ results in the 
condition~\cite{bulk1,chang}
\begin{equation}
J_0(x_n)\;Y_0(x_n e^{-kr\pi}) = J_0(x_n e^{-kr\pi})\;Y_0(x_n)~.
\end{equation}
which determines the KK masses.
For instance, for $e^{-kr\pi} = 10^{-16}$, we have $x_1\simeq 2.45$, $x_2\simeq 5.6$, 
$x_3\simeq 8.70$, $x_4\simeq 11.8$, $x_5\simeq 15.0$, etc. 
with the KK masses given by eqn.~(\ref{xndef}).  

The action for fermion fields in the bulk is given by~\cite{chang,gn}
\begin{equation}
S_f = \int d^4x~dy~ \sqrt{-g} \left\{ \frac{i}{2}\bar\Psi\hat{\gamma}^M
\left[{\cal D}_M
-\raisebox{0.12in}{$\leftarrow$}\hspace*{-0.18in}{\cal D}_M\right]\Psi   
- {\rm sgn}(y) M_f \bar\Psi\Psi\right\}~,
\label{sfermions} 
\end{equation}
where the covariant derivative in curved space is
\begin{equation}
{\cal D}_M \equiv \partial_M + \frac{1}{8}\,[\gamma^\alpha,~\gamma^\beta]~V_\alpha^N
~V_{\beta N;M}~,
\label{covder}
\end{equation} 
and $\hat{\gamma}^M\equiv V_\alpha^M\,\gamma^\alpha$, with 
$V_\alpha^M={\rm diag}(e^\sigma,e^\sigma,e^\sigma,e^\sigma,1)$ the inverse vierbein.
The bulk mass term $M_f$ in eqn.~(\ref{sfermions}) is expected to be of order
$k\simeq M_P$. 
Although the fermion field $\Psi$ is non-chiral, we can still define 
$\Psi_{L,R}\equiv \frac{1}{2}(1\mp\gamma_5)\Psi$. The KK decomposition can be written as
\begin{equation}
\Psi_{L,R}(x,y) = \frac{1}{\sqrt{2\pi r}}\,\sum_{n=0}\,\psi_n^{L,R}(x) e^{2\sigma} 
f_n^{L,R}(y)~,
\label{kkfermion}
\end{equation}
where $\psi_n^{L,R}(x)$ corresponds to the $nth$ KK fermion excitation and is 
a chiral four-dimensional field. 
Demanding that the KK fermions have the usual action in four dimensions leads to 
the coupled differential equations~\cite{chang,gn}
\begin{eqnarray}
(\partial_y - M_f)\,f_n^L(y) &=&-M_n\,e^\sigma\,f_n^R(y)\nonumber\\
(\partial_y + M_f)\,f_n^R(y) &=&~M_n\,e^\sigma\,f_n^L(y)~,
\label{diffeqfer}
\end{eqnarray}
where $M_n$ is the mass of the $nth$ KK fermion excitation. The corresponding 
normalization condition reads
\begin{equation}
\frac{1}{\pi r}\,\int_0^{\pi r}dy ~ e^\sigma\, f_n^{L,R}(y)\,f_m^{L,R}(y) 
= \delta_{nm}~,
\label{norfer}
\end{equation}
where we have made use of the fact that the $f_n(y)$ have definite $Z_2$ parity. 
The zero mode wave functions are obtained from eqn.~(\ref{diffeqfer})
for $M_n=0$. They are given by
\begin{equation}
f_0^{L,R}(y) = \sqrt{\frac{k\pi r\,(1\pm 2\nu_f)}{e^{k\pi r(1\pm2\nu_f)}-1}}\;
e^{\pm\nu_f \,k\,y}~,
\label{zeromode} 
\end{equation}
with $\nu_f\equiv M_f/k$ parametrizing the bulk fermion mass in units of the 
inverse AdS radius $k$. The $Z_2$ orbifold projection is used so that only 
one of these is actually allowed, either a left-handed or a right-handed zero mode.

The final piece of the bulk SM is the Higgs field. 
If it is allowed to live in the bulk, it gives a bulk mass term to the 
$W$ and $Z$ gauge bosons. In order to obtain the correct values of $M_W$ and $M_Z$
this bulk mass would have to be extremely fine tuned~\cite{chang}, in effect recovering the
same amount of tuning as in the SM.  
In order to avoid this problem, 
the Higgs field should be localized on the TeV boundary at 
$y=\pi r$. The picture of the five dimensional SM in the RS scenario
is particularly attractive when we take into account its potential to generate the 
hierarchy of fermion masses from ${\cal O}(1)$ flavor breaking in the fermion bulk 
mass parameter $\nu_f$. This was first considered in Ref.~\cite{gp} and further examined 
in \cite{huber1}. 
These authors have shown that allowing different values of $\nu_f$ 
within the natural constraint $|\nu_f|\simeq {\cal O}(1)$, results in exponentially
generated flavor hierarchies. For instance, one can generate 
the top quark mass with~\cite{gp,huber1} 
$\nu_t\simeq 0.5$,  and the electron mass with $\nu_e\simeq -0.5$. 

Other model building extensions, such as supersymmetry in the 
bulk~\cite{gp}, grand unification\cite{pom} and dynamical electroweak symmetry 
breaking~\cite{rius}
were considered, just to name a few. Thus, it is of great interest to study in detail 
what are the limitations of putting matter in the bulk.

Constraints on the bulk SM vary according to 
the localization of fermions in the fifth dimension. This is parametrized by the 
bulk mass parameter $\nu_f$. Very large positive values of $\nu_f$ 
correspond to fermions highly localized on the TeV boundary at $y=\pi r$, 
whereas negative $\nu_f$ corresponds to fermions with 
larger wave functions around the $y=0$ (Planck) boundary.
When fermions are localized in the TeV boundary they couple 
strongly to KK excitations of bulk gauge bosons, with the enhancement over 
the gauge coupling being given approximately by~\cite{bulk1}
$\simeq\sqrt{2\pi kr}\simeq 8.4$, resulting in a bound of $m_1>23$~TeV for the 
mass of the first  gauge boson excitation. 
This constraint is obtained from a fit to electroweak observables of the 
SM including the effect of KK excitations through the parameter $V$ defined in 
Ref.~\cite{bulk1} and arising from the tree-level exchange of the KK excitations.
However, when fermions are allowed to be in the bulk these bounds can be greatly relaxed.
This was noted in Refs.~\cite{chang,gp}, where the bound for the first KK mode of gauge
bosons is given as $m_1\simgt 2.1\left(g_1/g\right)$~TeV, with the 
ratio $(g_1/g)$ depending on 
the value of the bulk fermion mass parameter $\nu_f$. 
The localized fermion result is recovered for large positive $\nu_f$, 
$\nu_f=0$ gives $m_1\simgt 9$~TeV, and for negative 
values the bounds are much weaker, allowing the $1$~TeV mass range
\footnote{It should be 
noted that these bounds are on the mass of the first KK excitation of a gauge boson. 
But since
$m_1\simeq 2.45\Lambda_r$ (i.e. for $e^{k\pi r}=10^{16}$), the corresponding 
bounds on $\Lambda_r$ are weaker.}
In the next Section we will see that new non-oblique effects result in considerably 
stronger bounds.

\section{The effects of non-local wave-function renormalization}
\label{bounds}
In the presence of the Higgs VEV on the TeV boundary, the ``zero modes'' 
of the $W$ and $Z$ gauge bosons 
are no longer flat in the fifth dimension~\cite{huber1}. 
The resulting localized mass term repels the wave function in the vicinity of the
TeV boundary. 
In Ref.~\cite{cet} it was found that this 
leads to large tree level contributions to the oblique parameters $S$ and $T$. 
The $Z$ wave-function acquires a dependence on $y$ given by~\cite{cet} 
\begin{equation}
\chi^{(0)}_Z(y) \simeq  1 + \frac{M_Z^2}{4\Lambda_r^2}\left\{
2\pi kr -1 + (1-2ky)\,e^{2k(y-\pi r)}\right\}~,
\label{zwf} 
\end{equation}
where we have assumed $M_Z\ll\Lambda_r$, 
and the expression for the $W$ wave-function is obtained by replacing 
$M_Z\rightarrow M_W$. 
This leads to contributions to $S$ and $T$ which are approximately~\cite{cet} 
\begin{eqnarray}
S &\simeq & -4\pi \frac{v^2}{\Lambda_r^2}\,k\pi r~,\label{scont}\\
T &\simeq & -\frac{\pi}{2 c^2_w}\,\frac{v^2}{\Lambda_r^2}\,k\pi r~,\label{tcont}
\end{eqnarray}
where $v\simeq 246$~GeV, 
$c_w\equiv g/\sqrt{g^2+g'^2}$, with  $g,g'$ are the four-dimensional 
$SU(2)_L$ and $U(1)_Y$ gauge couplings respectively. 
In addition, the KK excitations of gauge bosons induce 
four-fermion interactions parameterized by a shift in $G_F$ given by~\cite{dhr1}
\begin{equation}
V = \sum_{n=1}^{\infty} \left(\frac{g_f^n(\nu)}{g_f^{\rm SM}}\right)^2
\,\frac{M_W^2}{m_n^2}~, 
\label{vdef}
\end{equation}
where $g_f^n$ denotes the coupling of the n-th gauge boson excitation to zero-mode
fermions, and $g_f^{\rm SM}$ the corresponding SM coupling. 
Since, unlike in Ref.~\cite{cet}, we are considering bulk fermions, their couplings
$g_f^n$ will depend on the bulk mass parameter $\nu$. The value of 
$V$ obtained in Ref.~\cite{cet} is recovered in the large positive $\nu$ limit. 
In that limit, all KK excitations couple with $g_f^n\simeq g_f^{\rm SM}\sqrt{2\pi k r}$. 
However, with fermions in the bulk and $|\nu|\simeq O(1)$, only 
the first KK excitation couples strongly, and eqn.~(\ref{vdef}) can be approximated
by its first term. This results in 
\begin{equation}
V \simeq \frac{g^2}{12}\,\frac{v^2}{\Lambda_r}\,k\pi r\;I^2(\nu)~,
\label{vcont}
\end{equation}
where we defined
\begin{equation}
I(\nu)\equiv \frac{1+2\nu}{1-e^{-k\pi r(1+2\nu)}}\,\int_0^1\,u^{1+2\nu}\,
\frac{J_1(x_1u) + \alpha_1\,Y_1(x_1u)}{|J_1(x_1) + \alpha_1\,Y_1(x_1)|}~,
\label{inu}
\end{equation}
with $\alpha$ given in eqn.~(\ref{alpha}) and $x_1$ defined by eqn.~(\ref{xndef}).
 
In Ref.~\cite{cet} the contributions from $S$, $T$ and $V$ are fit
to electroweak observables, in a scenario with only gauge fields propagating in the
bulk.
For  $m_h=115$~GeV they obtained the bound~\cite{cet} $\Lambda_r>11$~TeV at 
$95\%$ C.L. This constraint on the 
low energy scale $\Lambda_r$, translates into a lower bound on the 
lightest gauge boson KK excitation of $m_1>27$~TeV, and is slightly 
stronger than the ones previously obtained for fermions on the TeV brane and 
where only the $V$ parameter had been considered.  

Here we show that  when fermions are allowed in the bulk 
there are additional non-oblique effects in the couplings of $W$ 
and $Z$ to fermions. The electroweak gauge boson
wave-functions are normalized to their SM values at the low energy boundary. 
However, the couplings to fermions are a non-local quantity resulting from 
the overlap of the gauge boson and fermion wave-functions in the bulk. 
This effect is present even in the couplings to zero mode fermions, and is due to 
the $y$ dependence of the gauge boson wave-functions. 
In general the $5D$ coupling of fermions to a bulk
gauge boson ${\cal G}$ is given by 
\begin{equation}
S_{\rm int} = g_5\,\int d^4x \,dy\,\sqrt{-g}\, {\cal G}_\mu(x,y)\,\bar{\Psi}_f(x,y)
\hat{\gamma}^\mu \,\Psi_f(x,y)~,
\label{fdcoupling}
\end{equation} 
where $g_5= \sqrt{2\pi r}$, and $f$ is a fermion flavor index. 
This index has been kept since in principle it is possible that the 
fermion wave-functions~(\ref{zeromode}) are flavor dependent if 
the bulk mass parameter $\nu_f$ is not universal. The coupling 
of the gauge boson to a given fermion $f$ relative to its SM 
value is then given by
\begin{equation}
\left(\frac{g_f}{g^{\rm SM}_f}\right) = 
\frac{1}{\pi r}\,\int_0^{\pi r} dy \, e^{ky}\,|f_0^A(y)|^2\,\chi_G^{(0)}(y)~,
\label{goverg}
\end{equation}
with $A=(L,R)$ and $G=(W,Z)$. 
We define the new parameter 
\begin{equation}
\gamma_f^G \equiv \left(\frac{g_f}{g^{\rm SM}_f}\right) -1~.
\label{gammadef}
\end{equation}
where $f=b_L, b_R, t_L, \cdots$ is the zero mode fermion label.
From eqns.~(\ref{zeromode}) and~(\ref{zwf}) we have 
\begin{equation}
\gamma_f^Z = \frac{|f_0^A(0)|^2}{4k\pi r}\;\left(\frac{M_Z}{\Lambda_r}\right)^2\;
\left(I_1 + I_2 +I_3\right)~,
\label{gammaf}
\end{equation}
where we have defined 
\begin{eqnarray}
I_1&=& \frac{(2k\pi r-1)}{1\pm2\nu_f}\,\left[e^{(1\pm 2\nu_f)k\pi r} - 1 \right]~,
\nonumber\\
I_2 &=& \frac{e^{-2k\pi r}}{3\pm2\nu_f}\,\left[e^{(3\pm2\nu_f)k\pi r} - 1\right]~,
\label{theis}\\
I_3 &=& -\frac{2e^{-2k\pi r}}{3\pm2\nu_f}\,
\left\{ \left( k\pi r -\frac{1}{3\pm2\nu_f}\right)\,e^{(3+2\nu_F)k\pi r}
+\frac{1}{3\pm\nu_f}\right\}\nonumber~,
\end{eqnarray}
and the $+(-)$ corresponds to left-handed (right-handed) fermions.
A similar expression can be obtained for the shift in the W couplings
by noting that $\gamma_f^W=c_w^2\gamma_f^Z$.  
We first note that $\gamma_f^G$ always defines a {\em positive} shift in the 
corresponding coupling.  
We can also see that for $\nu_f<-0.5$,  the quantity $\gamma_f^G$ is practically independent
of $\nu_f$ and is given by
\begin{equation}
\gamma_f^Z \simeq \frac{M_Z^2}{4\Lambda_r^2}\,(2\pi kr-1)~.~~~~~~~~~~~~~~~(\nu_f<-0.5)
\label{gammapprox}
\end{equation}  
For larger values of $\nu_f$, the value of $\gamma_f^Z$ is reduced by the 
$\nu_f$ dependent terms. 

In order to study the effects of $\gamma_f^Z$ on the bounds on the scale $\Lambda_r$,  
we will first assume  that the bulk fermion mass is universal. 
In the next section we will study the effects of flavor violation in the bulk and the possible
signals associated with it.
We then consider, for the remaining of this section, the case where 
$\nu_{f_L}=-\nu_{f_R}\equiv\nu$ for all fermions propagating in the bulk.
With this choice, all couplings undergo a universal shift given by
\begin{equation} 
g_f \longrightarrow g_f\;(1+\gamma^G)~,
\label{gammauniv}
\end{equation}
where $G$ refers to either the $W$ or the $Z$ coupling.  Thus, there is a universal
shift in the charged current couplings such that 
\begin{equation}
\frac{G_F}{\sqrt{2}} = \frac{e^2}{8\,s^2_w\,c^2_w\,M_Z^2}\;(1+\gamma^W)^2~,
\label{gf}
\end{equation}
Following the standard procedure~\cite{peskin,burgess}, we redefine the Weinberg angle by
\begin{equation}
s_w\,c_w\;(1-\gamma^W)\longrightarrow  s_w\,c_w~. 
\label{redefsw}
\end{equation}
This means that we recover the standard form of the relation between $G_F$, $\alpha$, 
$s_w$ and $M_Z$, whereas the neutral current coupling now is 
\begin{equation}
\frac{e}{s_w\,c_w} \longrightarrow \frac{e}{s_w\,c_w}\;(1+\gamma^Z-\gamma^W)
 =\frac{e}{s_w\,c_w}\;(1+s^2_w\,\gamma^Z) ~.
\label{zcoupling}
\end{equation}
The replacement in eqn.~(\ref{redefsw}) implies that,
there is a new contribution to $s^2_w$ given by 
\begin{equation}
s^2_w\longrightarrow s^2_w(1+\cdots + \frac{c^2_w}{c^2_w-s^2_w}\,2\gamma^W)~,
\label{gwinsw2}
\end{equation}
where the dots stand for the contributions
from the $S$, $T$ and $V$ parameters. Eqn.~(\ref{gwinsw2}) implies an additional shift 
in fermion couplings to the $Z$ through the factor $(T_3^f-Q_f\,s^2_w)$.  

\begin{figure}
%
\centering
\epsfig{file=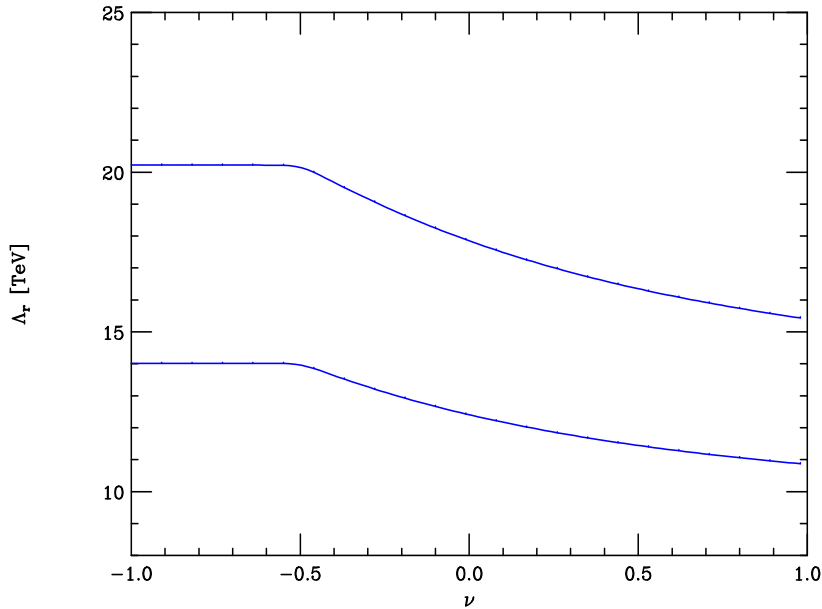,width=8cm,height=10.8cm,angle=90}
\caption{Lower bounds ($95\%$~C.L.) on $\Lambda_r$ vs. the fermion bulk mass parameter 
$\nu$.
The top curve corresponds to $m_h=115$~GeV, the lower curve is for  $m_h=300$~GeV.
}
\label{fig1}
\end{figure}

We perform a fit to the electroweak observables listed in the Appendix, where we also 
show the dependence on $S$, $T$, $V$ and $\gamma^Z$. The data is taken from Ref.~\cite{pdg}.
For a fixed value of the 
fermion bulk mass parameter $\nu$ we obtain bounds on the low energy scale
$\Lambda_r$. In Figure~\ref{fig1} we plot the $95\%$ C.L. bound on $\Lambda_r$ 
as a function of $\nu$. The top curve corresponds to $m_h=115$~GeV. 
We can see that the addition of the parameter $\gamma^Z$ arising from fermion
de-localization, 
results in stronger bounds.   
The constraint obtained in Ref.~\cite{cet} for fermions localized on the low energy
brane ($\Lambda_r>11$~TeV), is recovered in the $\nu \gg 1$ limit. 
It was pointed out in Refs.~\cite{gp,huber2} that the hierarchy of fermion
masses could be naturally obtained in the bulk SM for values of  
$\nu_f<-0.5$ for all fermions except the top quark. 
Since the flavor dependence of $V$ and $\gamma^Z$ is exponentially suppressed
for these values, the bounds in Figure~\ref{fig1} apply to this model. 
Thus the $95\%$~C.L. limit on $\Lambda_r$ in this scenario for generating
fermion masses is $\Lambda_r>20$~TeV, which translates into a 
mass bound for the first KK excitation of the gauge bosons which is 
$m_1>49$~TeV. 

Increasing the mass of the Higgs weakens the bounds on $\Lambda_r$ somehow, as it 
can be seen from the bottom curve in Figure~\ref{fig1}, for $m_h=300$~GeV. 
However, the quality of the fit worsens considerably as $m_h$ grows. 
The value of the $\chi^2$ is practically insensitive to the $\nu$ parameter. 
Varying then $\Lambda_r$ and $m_h$, the minimum $\chi^2$ is obtained for
the lighter $m_h$ and $\Delta\chi^2\approx 6.2$ for $m_h=300$~GeV. 
We then conclude that $m_h<300$~GeV at $95\%$~C.L. in the bulk SM of the 
Randall-Sundrum scenario. 

\begin{figure}
%
\centering
\epsfig{file=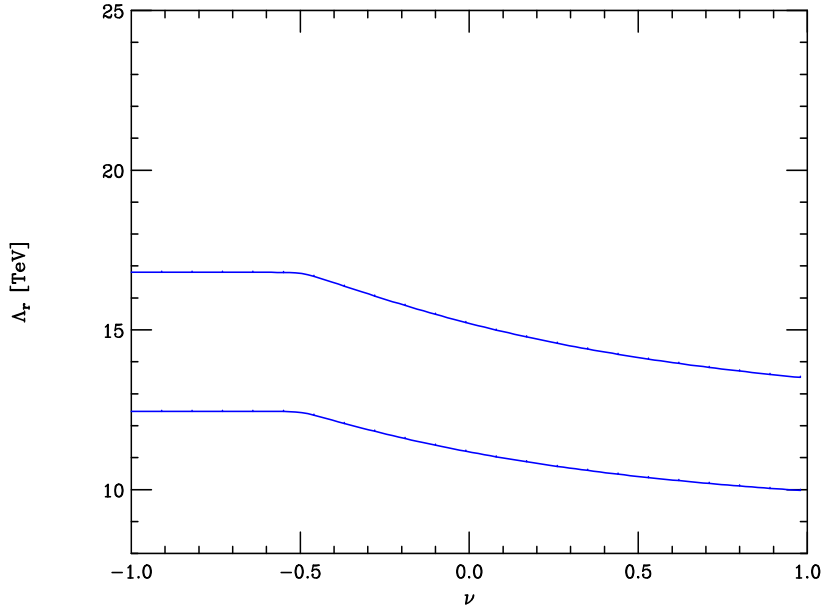,width=8cm,height=10.8cm,angle=90}
\caption{Lower bounds ($95\%$~C.L.) on $\Lambda_r$ vs. the fermion bulk mass parameter $\nu$,
for the case with the third generation confined to the low energy boundary.
The top curve corresponds to 
$m_h=115$~GeV, the lower curve is for  $m_h=300$~GeV.}
\label{fig2}
\end{figure}

We also study a scenario with the third generation  localized on the low
energy boundary.  It was recently suggested in Ref.~\cite{dhr2} that this 
is needed in order to avoid potentially large contributions to the $T$ parameter
from the KK excitations of the top quark. In Figure~\ref{fig2} we
plot the $95\%$~C.L. bounds on $\Lambda_r$, where the fermion 
bulk mass parameter $\nu$ now refers to that of the first two generations which live
in the bulk. Since the third generation does not propagate in the bulk, it does not
feels the effects of $\gamma^Z$. The resulting bounds are somewhat lower than the 
ones in Figure~\ref{fig1}. However, they are roughly three times stronger than the 
ones derived in \cite{dhr2}. For instance, for $\nu<-0.5$ Figure~\ref{fig2} implies
that the first gauge boson KK excitation must obey $m_1>41$~TeV, at $95\%$~C.L.
and for $m_h=115$~GeV. The lower curve corresponds to $m_h=300$~GeV and, just as 
for the case of Figure~\ref{fig1}, corresponds to $\Delta\chi^2\simeq 6.2$ and 
thus represents the $95\%$~C.L. value for $m_h$ in a fit of $\Lambda_r$ and 
$m_h$.   
 
We end this section with a comment on the possible effects of higher dimensional 
operators.
In principle, since the 5D theory is non-renormalizable we can 
write down higher-dimensional operators suppressed by the appropriate powers of the 
relevant 5D scale. As an example we consider the operator
\begin{equation}
\frac{c_1}{M_5}\,(\bar{\Psi}_f(x,y)\,\hat{\sigma}^{MN} \,\Psi_f(x,y))\,{\cal G}_{MN}(x,y)
\end{equation}
where $M_5\simeq M_P$ is the cut-off scale of the effective 5D theory and $c_1$ is a 
dimensionless coefficient which is naturally $c_1\simeq {\cal O}(1)$. 
Although this is suppressed with respect to the leading operator in 
eqn.~(\ref{fdcoupling}), 
it is possible that in reducing to the 4D effective theory the resulting operator may be 
suppressed only by the TeV scale due to the presence of the warp factor. 
For instance, projecting onto the zero modes results in the interaction
\begin{equation}
\frac{c_1}{M_5}\,\frac{1}{\sqrt{2\pi M_5 r}}\,\frac{(\frac{1}{2}+\nu)}{2(1+\nu)}\,
\frac{e^{2k\pi r(1+\nu)}-1}{e^{k\pi r(1+2\nu)}}
\left(\bar{f}_R\sigma^{\mu\nu} f_L\right)\,Z_{\mu\nu}~.
\label{fdhdop}
\end{equation}
This implies a contribution to the $Z$ couplings to the fermion $f$, which 
for the values of $\nu$ considered here gives 
\begin{equation}
\simeq  \frac{m_f}{M_Z}\,{\rm few}\times10^{-4}\,c_1~.
\label{shiftinzc}
\end{equation}
where we considered on-shell fermions. As usual, 
we impose $k<M_5/\sqrt{20}$ (resulting from asking that the curvature be smaller than 
$M_5^2$)
so there are no effects due to strong gravitational interactions.
Then, for the couplings we consider here 
this results in shifts that are typically $\simlt {\rm few}\times10^{-6}\,c_1$. This is 
at least two orders of magnitude smaller than the 
effects implied by the values of $\gamma^Z$ in eqns.~(\ref{gammaf}) 
and~(\ref{gammapprox}). We conclude that this particular operator can be safely
ignored. 
Nonetheless, we see that it is not suppressed by the Planck scale relative to the
leading operator in (\ref{fdcoupling}). 
This exercise highlights the fact that
the effects of higher dimensional operators are not {\em a priori} 
to be ignored and that,
in some cases, they could have important effects at the weak scale. 
However, for the purpose of our analysis, we assume these will not change considerably
the constraints derived in this section.

\section{Effects of Flavor Violation in the Bulk}
\label{flavor}
As it was mentioned in the previous section, it is possible to generate a large 
hierarchy in 
fermion masses if we allow ${\cal O}(1)$ flavor breaking in the bulk. Then we may allow  
the bulk fermion mass parameters $\nu_f$ to vary as long as they all 
are of order one (i.e. all bulk fermions
have masses of the order of $M_P$). This means that the shift in fermion 
couplings to the $W$ and $Z$
gauge bosons given in eqn.~(\ref{gammaf}) may be  non-universal. 
This necessarily leads  to flavor changing neutral currents (FCNC) of the $Z$ at tree level, 
as well as non-universal corrections to the charged current interactions. 
We concentrate here on the FCNC of the $Z$ due to their dangerous phenomenological nature. 
We first show how FCNC come about in the present context. 
Assuming that $\gamma_f^Z$ is different for each fermion induces a flavor dependent 
shift in the $Z$ coupling given by
\begin{equation}
\delta g_f = \gamma_f\,\frac{e}{s_w\,c_w}\,(T_3^f - Q_f\,s^2_w)~,
\label{delg}
\end{equation} 
where $T_3^f$ is the third component of weak isospin, $Q_f$ is the fermion 
electric charge and we
have dropped the superscript $Z$ in $\gamma_f$. 
The non-universality in these shifts results in FCNC when fermions are rotated from the 
weak to the mass eigenbasis.

Let us first consider the down quark sector. In the weak eigenbasis, the $Z$ coupling
to down quark types reads
\begin{eqnarray}
{\cal L} &=& - \frac{e}{s_w\,c_w}\,Z^\mu\,\left\{\left(-\frac{1}{2}+\frac{s^2_w}{3}\right)
\sum_{D=d,s,b} (1+\gamma_{D_L})(\bar D_L\gamma_\mu D_L) \right.\nonumber\\
&&\left.+\frac{s^2_w}{3}\,\sum_{D=d,s,b} (1+\gamma_{D_R})(\bar D_R\gamma_\mu D_R) 
\right\}~.
\label{zcwb}
\end{eqnarray} 
In principle, in eqn.~(\ref{zcwb}) there is also a factor of $(1-\gamma^W)$ coming from the
effect in charged currents, as in eqn.~(\ref{zcoupling}). However, this will be 
a flavor universal shift (as far as the quark flavor goes\footnote{Here we actually have
$(1-\frac{1}{2}\gamma^W_e -\frac{1}{2}\gamma^W_\mu)$ as entering in $\mu$ decay.}
). As we will see below, this 
will cancel in the FCNC effects since these 
will depend on differences of quark flavor-dependent quantities. 
We define the rotation of down quarks into the mass eigenbasis by 
$D_L\to A^L\;D_L$ and $D_R\to A^R\;D_R$ (here $D^T\equiv (d\,s\,b)$). 
There will be analogous rotation matrices in the 
up sector given by $U_L\to B^L\;U_L$, etc., such that $V_{CKM}=(B^L)^\dagger\,A^L$ is the
usual quark mixing matrix.
The unitarity of $A^L$ and $A^R$ implies that flavor off-diagonal terms not proportional
to a factor of $\gamma_D$ vanish. Then, off-diagonal terms are given by
\begin{eqnarray}
{\cal L}_D^{\rm FCNC} &=& -\frac{e}{s_w\,c_w}\,Z^\mu\,\left\{
\Delta_L^{ds}\,(\bar d_L\gamma_\mu s_L) + \Delta_L^{db}\,(\bar d_L\gamma_\mu b_L) 
+ \Delta_L^{sb}\,(\bar s_L\gamma_\mu b_L)
\right.\nonumber\\
&&\left.  + \;({\rm L}\longrightarrow {\rm R}) 
+ \;{\rm h.c.}
\right\}~,
\label{zcmb}
\end{eqnarray}
where we defined 
\begin{eqnarray}
\Delta_L^{ds}&\equiv & \gamma_{d_L}\,A_{11}^{L*}A_{12}^L + \gamma_{s_L}\,A_{21}^{L*}A_{22}^L
 + \gamma_{b_L}\,A_{31}^{L*}A_{32}^L~,
\label{dsl}\\
\Delta_L^{db}&\equiv & \gamma_{d_L}\,A_{11}^{L*}A_{13}^L + \gamma_{s_L}\,A_{21}^{L*}A_{23}^L
 + \gamma_{b_L}\,A_{31}^{L*}A_{33}^L~,
\label{dbl}\\
\Delta_L^{sb}&\equiv & \gamma_{d_L}\,A_{12}^{L*}A_{13}^L + \gamma_{s_L}\,A_{22}^{L*}A_{23}^L
 + \gamma_{b_L}\,A_{32}^{L*}A_{33}^L~.
\label{sbl}
\end{eqnarray}
 The analogous expressions for $\Delta^{ij}_R$ can be obtained by $L\to R$ in 
eqns.~(\ref{dsl})-(\ref{sbl}). Once again, we notice that if $\gamma_d=\gamma_s=\gamma_b$, 
the unitarity of $A^{L,R}$ implies that all off-diagonal terms vanish.
Unitarity also implies that eqns.~(\ref{dsl})-(\ref{sbl}) actually depend on two 
independent combinations of $\gamma_D$s, e.g. $(\gamma_{d_L}-\gamma_{s_L})$ and 
$(\gamma_{b_L}-\gamma_{s_L})$. Then, as mentioned above, any universal shift in the 
$Z$ couplings (\ref{zcwb}) cancels out when considering the flavor changing terms 
eqn.~(\ref{zcmb}).   
The interactions in eqns.~(\ref{dsl})-(\ref{sbl}) will induce $K^0-\bar{K^0}$, 
$B_d^0-\bar{B_d^0}$ and  $B_s^0-\bar{B_s^0}$ mixing, as well as rare $K$ and $B$ 
decays, all mediated by tree-level $Z$ exchange. 
Similar expressions can be written for the up quark sector. 

In principle, we have little information on the entries of $A^{L,R}$ or $B^{L,R}$. 
In order to illustrate how FCNC are generated let us examine a simple
model for the rotation matrices. Let us consider the situation where $B^L\simeq I$ and 
$A^L\simeq V_{\rm CKM}$. If this is the case then we have, for instance for the 
$d_L\to s_L$ term 
\begin{eqnarray}
\Delta^{ds}_L &\simeq& \gamma_{d_L}\,V_{ud}^*V_{us} +\gamma_{s_L}\,V_{cd}^*V_{cs}
+ \gamma_{b_L}\,V_{td}^*V_{ts} \nonumber\\
&\simeq& \lambda\,(\gamma_{d_L}-\gamma_{s_L})~,
\label{dslckm}
\end{eqnarray}
where the last line results from $V_{ud}^*V_{us} + V_{cd}^*V_{cs}\simeq 0$, and 
$\lambda\simeq0.22$ is the sine of the Cabibbo angle. 
We would also obtain
\begin{eqnarray}
\Delta^{db}_L &\simeq & (\gamma_{d_L}-\gamma_{s_L})\,V_{ud}^*V_{ub}
+ (\gamma_{b_L}-\gamma_{s_L})\,V_{td}^*V_{tb}~\nonumber\\ 
\Delta^{sb}_L &\simeq & (\gamma_{d_L}-\gamma_{s_L})\,V_{us}^*V_{ub}
+ (\gamma_{b_L}-\gamma_{s_L})\,V_{ts}^*V_{tb}~.
\label{dsbckm}
\end{eqnarray}
Although in general there is no reason to believe that $B^L\simeq I$, we 
will refer to this scenario (as well as to the general case) in order to evaluate 
the potential size of the effects in flavor physics.

We first consider FCNC processes in kaon physics induced by $Z$ exchange.
The contribution to $K^0-\bar{K^0}$ mixing is given by
\begin{eqnarray}
\Delta m_K &=& \frac{4\,G_F}{\sqrt{2}}\,f_K^2\,m_K \,
\left\{ \frac{2}{3}\,(-\frac{1}{2}\frac{s^2_w}{3})^2\,Re[(\Delta^{ds}_L)^2]
+\frac{2}{3}\,(\frac{s^2_w}{3})^2\,Re[(\Delta^{ds}_R)^2] \right.\nonumber\\
&&\left.-4(-\frac{1}{2}+\frac{s^2_w}{3})(\frac{s^2_w}{3})\,
\left[\frac{1}{4}+\frac{1}{6}\left(\frac{m_K}{m_s+m_d}\right)^2\right]\,
Re[\Delta^{ds}_L\,\Delta^{ds}_R]
\right\}~,
\label{dmk}
\end{eqnarray}
Even in the presence of the last term, which  is ``chirally enhanced'' and dominates, 
we do not find a large effect.
If we assume that the $Z$ exchange contribution saturates the experimental 
measurement~\cite{pdg}
$\Delta m_K^{\rm exp}=(3.489\pm0.008)\times10^{-15}$~GeV, then considering
$\Delta^{ds}_L\simeq\Delta^{ds}_R$ , we find 
$\sqrt{Re[(\Delta^{ds}_L)^2]}\;<\;1\times10^{-4}$. 

A tighter bound is obtained from 
$K_L\to\mu^+\mu^-$. This is extracted from~\cite{burgess}
\begin{eqnarray}
\frac{Br(K_L\to\mu^+\mu^-)}{Br(K^+\to\mu^+\nu_\mu)}
&=&\frac{\tau(K_L)}{\tau(K^+)}\,\frac{8}{|V_{us}|^2}\,
\left[(-\frac{1}{2}+s^2_w)^2+(s^2_w)^2\right]\nonumber\\
&&\times\left[(-\frac{1}{2}+\frac{s^2_w}{3})^2\,|\Delta^{ds}_L|^2
+(s^2_w)^2\,|\Delta^{ds}_R|^2\right]~.
\label{kmmbound}
\end{eqnarray}
With~\cite{pdg} $Br(K_L\to\mu^+\mu^-)=(7.18\pm0.17)\times10^{-9}$ and assuming that
the new contribution saturates the rate, we obtain 
$\Delta^{ds}_L\;<\;4.8\times10^{-5}$. 

But the best bound
on $\Delta^{ds}_{L,R}$ comes from $K^+\to\pi^+\nu\bar\nu$. The contribution 
to the amplitude is given by
\begin{equation}
\delta{\cal A} = \frac{e^2}{s^2_w\,c^2_w\,M_Z^2}\,\frac{1}{2}\,
\left\{ \left(-\frac{1}{2}+\frac{s^2_w}{3}\right)\,\Delta^{ds}_L (\bar d_L\gamma_\mu s_L) +
\frac{s^2_w}{3}\,\Delta^{ds}_R\,(\bar d_R\gamma_\mu s_R)
\right\}\times\sum_{i=e,\mu,\tau} (\bar\nu^i_L\gamma^\mu\nu^i_L) + {\rm h.c.}~.
\label{kpnnnew}
\end{equation}
On the other hand, the SM amplitude can be written as~\cite{bbl} 
\begin{equation}
{\cal A}_{\rm SM} = \frac{G_F}{\sqrt{2}}\,\frac{\alpha}{\pi\,s^2_w}\,S\,
(\bar d_L\gamma_\mu s_L)\times\sum_{i=e,\mu,\tau} (\bar\nu^i_L\gamma^\mu\nu^i_L) 
+ {\rm h.c.}~,
\label{kpnnsm}
\end{equation}
where $S\simeq 2\times10^{-3}$. The current experimental measurement is~\cite{e787}
$Br(K^+\to\pi^+\nu\bar\nu)=(1.57^{+1.75}_{-0.82})\times10^{-10}$, or 
an upper bound of $Br(K^+\to\pi^+\nu\bar\nu)<5\times10^{-10}$. The left-handed 
term in (\ref{kpnnnew}) interferes with the SM amplitude. If we consider this 
term only, we derive the $2\sigma$ bound
\begin{equation}
\Delta^{ds}_L < 1.2\times10^{-5}~.
\label{dslbound}
\end{equation}
Conversely, if we only consider the right-handed contribution we obtain
\begin{equation}
\Delta^{ds}_R < 1.9\times10^{-5}~.
\label{dsrbound}
\end{equation}
Finally, if we consider $\Delta^{ds}_L\simeq\Delta^{ds}_R$, then the bound is
\begin{equation}
\Delta^{ds}_{L,R} < 7.5\times10^{-6}~.
\label{dslrbound}
\end{equation}
In order to estimate the compatibility of these bounds with the 
ones derived in the previous section from electroweak precision measurements, 
we assume $B\simeq I$ so we can make use of eqn.~(\ref{dslckm}) which, together 
with the bound in (\ref{dslbound}) gives
\begin{equation}
(\gamma_{d_L}-\gamma_{s_L})< 5.5\times10^{-5}~.
\label{gdsbound}
\end{equation}
If we now consider as reference the bounds on $\Lambda_r$ that we obtained in the 
previous section and remember that for $\nu_f<-0.5$ $\gamma_f$ is $\nu_f$-independent, 
we can derive bounds on bulk mass differences. For instance, for 
$\Lambda_r>20~$TeV, eqn.~(\ref{gdsbound}) implies $(\nu_d-\nu_s)<0.1$, 
as long as one 
of them is $>-0.5$. For $\Lambda_r>13~$TeV, $(\nu_d-\nu_s)<0.08$, 
for $\Lambda_r>10$~TeV, $(\nu_d-\nu_s)<0.05$, etc.  
Then, although the bounds imply a certain amount of tuning between the bulk mass 
parameters, this is not particularly worrisome. If both bulk mass parameters are
$(\nu_d, \nu_s) <-0.5$, the bound (\ref{gdsbound}) is easily satisfied. 
In the most general case, there is too much freedom in order to 
use eqns.~(\ref{dslbound}) and~(\ref{dsrbound}) to constraint $\Lambda_r$ and
the bulk mass parameters. This is particularly true in the case of $\Delta^{ds}_R$, 
since -even  assuming that $A^{L}\simeq V_{\rm CKM}$ holds- we have no information
on $A^R$. However, the lesson we draw from the bound in eqn.~(\ref{gdsbound}) is that
$K^+\to\pi^+\nu\bar\nu$ is sensitive to small flavor breaking in the bulk, 
particularly taking into account future improvements in the experimental measurements 
of this mode as well as $K_L\to\pi^0\nu\bar\nu$.

Observables in $B$ physics turn out to be less sensitive. For instance, 
following the derivation 
of eqn.~({\ref{dmk}), $B^0-\bar{B^0}$ mixing results in the bound 
$\Delta^{db}_{L,R}<5\times10^{-4}$, where again we assumed $\Delta^{db}_L\simeq
\Delta^{db}_R$. 
This in general corresponds to a much weaker constraint on $(\gamma_{d_L}
-\gamma_{b_L})$, as it is illustrated by the scenario with $A^L\simeq V_{\rm CKM}$. 
In this case, $\Delta^{db}_L\simeq \lambda^3\,(\gamma_{d_L}-\gamma_{b_L})$, so that
the effect is largely suppressed by the large power of the Cabibbo angle.
The bound gets weaker if we assume only one of them 
non-vanishing, since the last term in (\ref{dmk}) is still very important (although not
dominant). Although $\Delta^{db}_R$ could be large and unsuppressed by powers of 
$\lambda$, its effect in $B^0-\bar{B^0}$ mixing is suppressed -relative to that of 
$\Delta^{sb}_L$- by 
two powers of $(s^2_w/3)/(-1/2+s^2_w)\simeq 0.18$. 
The current bounds from rare semileptonic 
$B$ decays such as $b\to s\ell^+\ell^-$ are even weaker. Although a great deal of 
improvement in the experimental situation is expected soon in these decay modes, these
are still not sensitive to $\Delta^{sb}_{L,R}$ once we consider the bounds
on $\Lambda_r$ from the previous section.

Finally, we consider the effects of flavor violation in the up quark sector.
There is an expression analogous to eqn.~(\ref{zcwb}) for the $Z$ couplings 
to up quarks, which can be obtained by replacing $D$ by $U=(u~c~t)^T$, and 
by putting the appropriate factors of $(T_3-Q_fs^2_w)$. 
There are also expressions for $\Delta^{uc}_{L,R}$, $\Delta^{ut}_{L,R}$
and $\Delta^{ct}_{L,R}$ similar to those in eqns.~(\ref{dsl})-(\ref{sbl}), 
where the rotation matrix is now $B^{L,R}$.  
The terms involving the top quark will only surface in top
rare decays, which require top data samples not yet available. 
We will then concentrate on the  $uc$ terms. 
Current experimental constraints on $D^0-\bar{D^0}$ mixing~\cite{ddmix}
result in\footnote{This bound is obtained 
assuming that there is no strong relative phase between
the doubly Cabibbo suppressed decay $D^0\to K^+\pi^-$ and the 
Cabibbo allowed mode $D^0\to K^-\pi^+$.} 
$\Delta m_D<4.5\times10^{-14}$~GeV.  The contributions of $\Delta^{uc}_{L,R}$ 
to $\Delta m_D$ can be read off eqn.~(\ref{dmk}). If we assume 
$\Delta^{uc}_L\simeq \Delta^{uc}_R$ we obtain
\begin{equation}
\Delta^{uc}_{L,R} < 3\times10^{-4}~.
\label{leqrbound}
\end{equation}
Unlike in the case of $\Delta^{db}_{L,R}$, here we do not expect a large suppression 
by powers of $\lambda$. For instance, if we assume that all of the CKM matrix comes from 
the up sector, i.e. $B^{L}\simeq V_{\rm CKM}$ and $A^{L}\simeq I$, then 
$\Delta^{uc}_L\simeq \lambda\,(\gamma_{u_L}-\gamma_{c_L})$, only suppressed by $\lambda$. 
Moreover, it is possible to imagine that $\Delta^{uc}_R$ is unsuppressed, which would 
mean that $D^0-\bar{D^0}$ mixing is sensitive to   
$(\gamma_{u_L}-\gamma_{c_L})\simeq 3\times10^{-4}$. 
If we assume that only $\Delta^{uc}_L$ is non-vanishing, we obtain
$\Delta^{uc}_L<5\times10^{-4}$. If, on the other hand, we assume that only the right-handed
term is present, we obtain  $\Delta^{uc}_R<5\times10^{-3}$. 
From the bounds derived
in the previous section we know that $\gamma_f\simlt 1\times10^{-3}$ for $\nu_f<-0.5$, and that 
the flavor dependence reduces $\gamma_f$ as $\nu_f$ becomes more positive.
We then conclude that the flavor universal bounds from 
electroweak precision measurements are compatible with the bounds on $\Delta^{uc}_{L,R}$
from $\Delta m_D$, even for relatively large values of $(\nu_u-\nu_c)$, the difference in the 
relevant bulk mass parameters. Future improvements in the experimental 
bound on $\Delta m_D$ will begin to probe flavor violation in the bulk 
in a region where the flavor dependence in $\gamma_f$ could manifest itself.
However, at the moment this is not a constraint comparable to the ones derived above 
from kaon processes. 
On the other hand, rare $D$ decays receive very small contributions from 
$\Delta^{uc}_{L,R}$ once the bounds derived from $\Delta m_D$ are taken into account.
For instance, in $D\to (\pi,\rho)\ell^+\ell^-$, they mainly contribute to 
the four-fermion operator $(\bar{u}_L\gamma_\mu c_L)(\bar\ell\gamma^\mu\gamma_5\ell)$.
Even when looking at the low dilepton
mass region, away from the dominating resonant contributions, the 
effect of $\Delta^{uc}_{L,R}$ are small enough to be comparable to the 
remaining hadronic uncertainties present~\cite{bghp}.

\section{Conclusions}
\label{conclusions}
We have derived constraints on the low energy scale $\Lambda_r$ induced in the 
RS scenario, when the SM fields are allowed to propagate in the 5D bulk. 
The effects are the result of the deformation of the $Z$ and $W$ zero-mode 
wave-functions due to the presence of the Higgs filed in the low energy boundary. 
When only gauge fields are allowed in the bulk, this  only leads to the contributions
to the oblique parameters $S$ and $T$ discussed in Ref.~\cite{cet}. If fermions 
also propagate in the bulk, non-oblique effects arise due to the modified 
couplings of the $Z$ and the $W$ to {\em zero mode } fermions. We considered these new 
effects here. 
In Section~\ref{bounds} we studied the constraints one obtains by 
assuming 
that the shifts in the couplings to fermions are flavor universal. 
This is a rather good assumption, as we point out in the discussion leading to 
eqn.~(\ref{gammapprox}): for values of the fermion bulk mass parameter $\nu_f<-0.5$, 
corrections to it are exponentially suppressed.
This is the region of $\nu_f$ that can be used in order to generate the fermion masses
as pointed out in Refs.~\cite{gp,huber2}.
For $\nu_f>-0.5$, the dependence of the bounds on $\nu_f$ is rather mild. 
Since the mass of the first KK excitation of a gauge boson is given by 
$m_1\simeq 2.45\,\Lambda_r$, we conclude that these must be heavier than 
a few tens of TeV, putting them out of reach of the Large Hadron Collider at CERN.

We also found -just as in Ref.~\cite{cet} for the case of
localized fermions- that although the bounds on $\Lambda_r$ 
are somewhat relaxed as the Higgs mass increases, the $\chi^2$ considerably worsens.
We observe the in the bulk SM $m_h<300$~GeV at $95\%$~C.L., a bound very similar
to the one derived in \cite{cet}.

We also considered the scenario motivated in Ref.~\cite{dhr2}, where the 
third generation is confined to the low energy boundary. The bounds on $\Lambda_r$
are displayed in Figure~\ref{fig2}. Although these are somehow lower than those in 
Figure~\ref{fig1}, they are still about tree times stronger than the ones obtained in 
Ref.~\cite{dhr2}. 

In Section~\ref{flavor}, we considered the residual effects of flavor dependence
in $\gamma^Z_f$ by allowing ${\cal O}(1)$ flavor breaking in the 
fermion bulk mass parameters $\nu_f$. This induces FCNC of the $Z$ to zero-mode
fermions. We showed that these FCNC effects are not dangerous once the 
constraints on $\Lambda_r$ derived in Section~\ref{bounds} are taken into account.
We found the  most sensitive observables to be 
in kaon physics such as $K^0-\bar{K^0}$ mixing, 
$K_L\to \mu^+\mu^-$, but especially $K^+\to\pi^+\nu\bar\nu$. 
The constraints from these imply that there should be a certain amount of degeneracy in 
bulk masses of the down quark sector. Although the bounds do not result in 
fine tuning, better measurements of $K^+\to\pi^+\nu\bar\nu$ may 
imply the need of a higher
$\Lambda_r$ if one is to avoid $1\%$ adjustments in the bulk masses.   

We stress that the effects considered here occur at tree-level and are the result 
of interactions among zero-mode gauge bosons and fermions.   
Additional contributions to oblique and non-oblique parameters may result from 
loop effects involving KK excitations. However, in most cases we do not expect the 
predictions to be well defined. In particular, we expect that cut-off
dependence would hinder our ability to translate a one loop calculation into a 
bound on the parameters of the theory. This sensitivity to the cut-off 
can be interpreted 
as a consequence of the fact that the 5D theory is non-renormalizable, so 
higher-dimensional operators with unknown coefficients may absorb the cut-off dependence
from a naive loop computation. 
As we show at the end of Section~\ref{bounds}, 
the effect of higher-dimensional 
operators could partially cancel some of the effects discussed here. 
However, it appears unnatural to expect that they could  do so 
efficiently enough to loosen the constraint considerably. 

There are also flavor violating effects induced by the mixing of zero-mode fermions 
with KK excitations~\cite{aguila}. However, these become irrelevant once $\Lambda_r$ 
is raised to the values shown in Figures~\ref{fig1} and~\ref{fig2}.

\vskip1.0cm
\noindent
{\bf Acknowledgments}
The author thanks Mike Chanowitz and John Terning for useful comments,
and especially 
Walter  Goldberger for useful discussions and suggestions during the course of this work.
This work was supported by the Director, Office of Science, 
Office of High Energy and Nuclear Physics of the U.S. Department of 
Energy under Contract DE-AC0376SF00098.

\newpage
\appendix
\section*{Appendix}
\setcounter{equation}{0}
\renewcommand{\theequation}{A.\arabic{equation}}

What follows are the expressions of observables used in the fits of 
Section~\ref{bounds} as functions of $S$, $T$, $V$ and $\gamma^Z$. 

\begin{eqnarray}
&&\Gamma_Z = (\Gamma_Z)^{\rm SM}\left(1-3.8\times10^{-3}S+0.011T -1.4V-0.08\gamma^Z\right)
\nonumber\\
&&R_e = (R_e)^{\rm SM}\left(1-2.9\times10^{-3}S+2\times10^{-3}T-0.26V-0.4\gamma^Z\right)
\nonumber\\
&&R_\mu = (R_\mu)^{\rm SM}\left(1-2.9\times10^{-3}S+2\times10^{-3}T-0.26V-0.4\gamma^Z\right)
\nonumber\\
&&R_\tau = (R_\tau)^{\rm SM}\left(1-2.9\times10^{-3}S+2\times10^{-3}T-0.26V-0.4\gamma^Z\right)
\nonumber\\
&&\sigma_h = (\sigma_h)^{\rm SM}\left(1+2.2\times10^{-4}S-1.6\times10^{-4}T
-0.021V-0.96\gamma^Z\right)
\nonumber\\
&&\Gamma_b = (\Gamma_b)^{\rm SM}\left(1-4.5\times10^{-3}S+0.011T-1.4V-0.18\gamma^Z\right)
\nonumber\\
&&\Gamma_c = (\Gamma_c)^{\rm SM}\left(1-6.5\times10^{-3}S+0.0124T-1.6V+0.45\gamma^Z\right)
\nonumber\\
&&\Gamma_{\rm inv.} =(\Gamma_{\rm inv.})^{\rm SM}\left(1+7.8\times10^{-3}T-V+0.46\gamma^Z\right)
\nonumber\\
&&A^e_{\rm FB} = (A^e_{\rm FB})^{\rm SM}-6.8\times10^{-3}S+4.8\times10^{-3}T-0.62V-0.95\gamma^Z
\nonumber\\
&&A^\mu_{\rm FB} = (A^\mu_{\rm FB})^{\rm SM}-6.8\times10^{-3}S+4.8\times10^{-3}T-0.62V
-0.95\gamma^Z \nonumber\\
&&A^\tau_{\rm FB} = (A^\tau_{\rm FB})^{\rm SM}-6.8\times10^{-3}S+4.8\times10^{-3}T-0.62V
-0.95\gamma^Z \nonumber\\
&&A_\tau(P_\tau) = (A_\tau(P_\tau))^{\rm SM} -0.028S+0.020T-2.6V-4\gamma^Z
\nonumber\\
&&A_e(P_\tau) = (A_e(P_\tau))^{\rm SM} -0.028S+0.020T-2.6V-4\gamma^Z
\nonumber\\
&&A^b_{\rm FB} = (A^b_{\rm FB})^{\rm SM} -0.020S+0.014T-1.8V-2.77\gamma^Z
\nonumber\\
&&A^c_{\rm FB} = (A^c_{\rm FB})^{\rm SM} -0.016S+0.011T-1.4V-2.16\gamma^Z
\nonumber\\
&&A_{\rm LR} = (A_{\rm LR})^{\rm SM} -0.028S+0.02T-2.6V -4\gamma^z
\nonumber\\
&&M_W = (M_W)^{\rm SM}\left(1-3.6\times10^{-3}S+5.5\times10^{-3}T-0.71V+0.66\gamma^Z\right)
\nonumber\\
&&g^2_L(\nu N\to\nu N)=(g^2_L(\nu N\to\nu N))^{\rm SM}-2.7\times10^{-3}S+6.5\times10^{-3}T
-0.066V-0.096\gamma^Z
\nonumber\\
&&g^2_R(\nu N\to\nu N)=(g^2_R(\nu N\to\nu N))^{\rm SM}+9.3\times10^{-4}S+2.0\times10^{-4}T
+0.1V+0.16\gamma^Z
\nonumber\\
&&g_{eV}(\nu e\to\nu e)=( g_{eV}(\nu e\to\nu e))^{\rm SM}+7.2\times10^{-3}S-5.4\times10^{-3}T
+0.65V+0.99\gamma^Z
\nonumber\\
&&g_{eA}(\nu e\to\nu e)=( g_{eA}(\nu e\to\nu e))^{\rm SM}-3.9\times10^{-3}T
+0.15V-0.23\gamma^Z
\nonumber\\
&&Q_W(Cs) = (Q_W(Cs))^{\rm SM} -0.793S-0.009T-0.47V-290.8\gamma^Z
\end{eqnarray}


\begin{thebibliography}{99}

\bibitem{nima1}
N.~Arkani-Hamed, S.~Dimopoulos and G.~R.~Dvali,
Phys.\ Lett.\ B {\bf 429}, 263 (1998); 
I.~Antoniadis, N.~Arkani-Hamed, S.~Dimopoulos and G.~R.~Dvali,
Phys.\ Lett.\ B {\bf 436}, 257 (1998).
%
\bibitem{rs1}
L.~Randall and R.~Sundrum,
Phys.\ Rev.\ Lett.\  {\bf 83}, 3370 (1999).
%
\bibitem{bulk1} W.~D.~Goldberger and M.~B.~Wise,
Phys.\ Rev.\ D {\bf 60}, 107505 (1999); 
H.~Davoudiasl, J.~L.~Hewett and T.~G.~Rizzo,
Phys.\ Lett.\ B {\bf 473}, 43 (2000); 
A.~Pomarol, Phys.\ Lett.\ B {\bf 486}, 153 (2000); 
%
\bibitem{chang} 
S.~Chang, J.~Hisano, H.~Nakano, N.~Okada and M.~Yamaguchi,
Phys.\ Rev.\ D {\bf 62}, 084025 (2000). 
%
\bibitem{gn}
Y.~Grossman and M.~Neubert,
Phys.\ Lett.\ B {\bf 474}, 361 (2000).
%
\bibitem{gp}T.~Gherghetta and A.~Pomarol,
Nucl.\ Phys.\ B {\bf 586}, 141 (2000).  
%
\bibitem{huber1} S.~J.~Huber and Q.~Shafi,
Phys.\ Rev.\ D {\bf 63}, 045010 (2001). 
%
\bibitem{dhr1} H.~Davoudiasl, J.~L.~Hewett and T.~G.~Rizzo,
Phys.\ Rev.\ D {\bf 63}, 075004 (2001).
%
\bibitem{dhr2}
J.~L.~Hewett, F.~J.~Petriello and T.~G.~Rizzo,
{\em ``Precision measurements and fermion geography in the Randall-Sundrum  
model revisited''}, {\bf hep-ph/0203091}.
%
\bibitem{cet} C.~Cs\'{a}ki, J.~Erlich and J.~Terning,
{\em ``The effective Lagrangian in the Randall-Sundrum model and electroweak  physics''}, 
{\bf hep-ph/0203034}.
%
\bibitem{pom}A.~Pomarol,
Phys.\ Rev.\ Lett.\  {\bf 85}, 4004 (2000);
L.~Randall and M.~D.~Schwartz,
JHEP {\bf 0111}, 003 (2001); L.~Randall and M.~D.~Schwartz,
Phys.\ Rev.\ Lett.\  {\bf 88}, 081801 (2002); 
W.~D.~Goldberger and I.~Z.~Rothstein,
{\em``High energy field theory in truncated AdS backgrounds''}
{\bf hep-th/0204160}.
%
\bibitem{rius}N.~Rius and V.~Sanz,
Phys.\ Rev.\ D {\bf 64}, 075006 (2001).
%
\bibitem{peskin} M.~E.~Peskin and T.~Takeuchi,
Phys.\ Rev.\ Lett.\  {\bf 65}, 964 (1990).
\bibitem{burgess}
C.~P.~Burgess, S.~Godfrey, H.~Konig, D.~London and I.~Maksymyk,
Phys.\ Rev.\ D {\bf 49}, 6115 (1994).
%
\bibitem{pdg}
D.~E.~Groom {\it et al.}  [Particle Data Group Collaboration],
Eur.\ Phys.\ J.\ C {\bf 15}, 1 (2000); and updates from http://www-pdg.lbl.gov.
%
\bibitem{huber2}S.~J.~Huber and Q.~Shafi,
Phys.\ Lett.\ B {\bf 498}, 256 (2001). 
%
\bibitem{bbl}G.~Buchalla, A.~J.~Buras and M.~E.~Lautenbacher,
Rev.\ Mod.\ Phys.\  {\bf 68}, 1125 (1996). 
%
\bibitem{e787}
S.~Adler {\it et al.}  [E787 Collaboration],
Phys.\ Rev.\ Lett.\  {\bf 88}, 041803 (2002); 
S.~Adler {\it et al.}  [E787 Collaboration],
{\em ``Search for the decay $K^+ \to\pi^+\nu\bar\nu$ in the momentum region  
$P(\pi) < 195~$MeV''}
{\bf hep-ex/0201037}.
%
\bibitem{ddmix} G.~Brandenburg {\it et al.}  [CLEO Collaboration],
Phys.\ Rev.\ Lett.\  {\bf 87}, 071802 (2001).
%
\bibitem{bghp} G.~Burdman, E.~Golowich, J.~Hewett and S.~Pakvasa,
{\em ``Rare charm decays in the standard model and beyond''}, 
{\bf hep-ph/0112235}.
%
\bibitem{aguila} 
F.~del Aguila and J.~Santiago,
Phys.\ Lett.\ B {\bf 493}, 175 (2000).
%
\end{thebibliography}
\end{document}